# Lower bound of computational complexity of knapsack problems

Zhidong Zhang*

Shenyang National Laboratory for Materials Science, Institute of Metal Research, Chinese Academy of Sciences, Shenyang 110016, China

**\* Correspondence:** Email: zdzhang@imr.ac.cn.

**Abstract:** The quantum statistics mechanism is very powerful for investigating the equilibrium states and the phase transitions in complex spin disorder systems. The spin disorder systems act as an interdisciplinary platform for solving the optimum processes in computer science. In this work, I determined the lower bound of the computational complexity of knapsack problems. I investigated the origin of nontrivial topological structures in these hard problems. It was uncovered that the nontrivial topological structures arise from the contradictory between the three-dimensional character of the lattice and the two-dimensional character of the transfer matrices used in the quantum statistics mechanism. I illustrated a phase diagram for the non-deterministic polynomial (NP) vs polynomial (P) problems, in which a NP-intermediate (NPI) area exists between the NP-complete problems and the P-problems, while the absolute minimum core model is at the border between the NPI and the NP-complete problems. The absolute minimum core model of the knapsack problem cannot collapse directly into the P-problem. Under the guide of the results, one may develop the best algorithms for solving various optimum problems in the shortest time (improved greatly from $O(1.3^N)$ to $O((1 + \varepsilon)^N)$ with $\varepsilon \to 0$ and $\varepsilon \neq 1/N$) being in subexponential and superpolynomial. This work illuminates the road on various fields of science ranging from physics to biology to finances, and to information technologies.

## 1. Introduction

In recent years, there has been great progresses in computer science, especially in machine learning, artificial intelligence, and so on. It is well known that neural networks for artificial intelligence are closely related with spin-glass models in statistic physics. The study of the computational complexity of a complex problem, for calculations of physical properties of a complicated system, like a spin disorder system, is an extremely important topic either in physics, mathematics or in computer science. On the one hand, one would develop the optimum algorithm to reveal the physical properties of a complicate system by guide of the lower bound of the computational complexity. The optimum algorithm can find the solution of the spin disorder system (or related physical systems) in the shortest time, with the sufficient accuracy and within the high precision. On the other hand, one would develop the optimum processes to investigate various hard problems in mathematics and computer science, under the guide of physical insight obtained from solving the spin disorder system.

In physics, it is always a main target to inspect the ground state of a complex system and calculate its thermodynamic functions (such as internal energy, enthalpy, entropy, free energy, etc.) [1]. From the differential of the free energy, we can determine the thermodynamic properties; for instance, specific heat, spontaneous magnetization, susceptibility, correlation functions, and correlation length. The spin disorder systems are among complex systems, in which abundant physical phenomena may exist and the processes for calculations of physical properties may be very complicated. It is important to find out analytically the solution of a physical system for having a deep understanding. Unfortunately, until now, we have known the exact solutions of only few physical systems. For the complicated systems, like the spin disorder systems, it is extremely difficult to figure out the exact solution, so researchers usually develop algorithms to numerically simulate the exact solution. For this purpose, researchers attempt to design the optimum algorithm to find/reach the exact solution with sufficient accuracy and within the high precision in the shortest time.

In magnetic systems, there are several sources that introduce disorders. The temperature, or the thermo-activity, causes the disorder phase at high temperatures. The disorder can be introduced initially in Hamiltonian of a spin glass system, by terms of diluted spins with random distributions, or disorderly distributed competing interactions between spins. In the former case, the spins at different lattice sites may have different values, whereas in the latter case, the interactions between spins may have different strengths and signs. The introduction of these disorder terms in the Hamiltonian results in rich magnetic states and abundant physical phenomena [2–4], and the study of their complexity is important for developing efficient algorithms for optimum problems [5,6]. Moreover, the introduction of the disorder terms makes the computational processes for the thermodynamic properties much more complicated. The Ising model is a well-known model widely used for description of phase transitions and applied as a paradigm for various complex systems [7–9]. The spin-glass Ising model is a generalization of the ferromagnetic Ising model [4]. The calculation of the ground state (and other equilibrium physical properties) of the three-dimensional (3D) spin-glass Ising model $M_{SGI}^{3D}$ [10,11] has been proven to be one among the NP-complete problems (Cook- Levin Theorem [12,13]).

In another study [14], I analyzed characters of the 3D spin-glass Ising model, such as topological effect, randomness, frustration, and non-ergodic behavior, to show its nature of a NP-complete problem. In previous work [15], I investigated the mapping between the 3D spin-glass Ising model and the Boolean satisfiability (K-SAT) problems. The major reasons that the spin glass model was chosen as a research model for NP-complete problems were because: 1) The 3D spin-glass Ising model is a NP-complete system with the existence of randomness and nonlocality, causing the non-triviality similar to other NP-complete problems, while the two-dimensional (2D) spin-glass Ising model is a P problem. 2) With the Clifford algebraic representation, it is easy to reveal nontrivial topological structures, non-planarity graphs, nonlocalities, or long-range spin entanglements in the 3D spin-glass Ising model. 3) It is the easiest system to figure out the absolute minimum core (AMC) model for computational complexity among all NP-complete problems, since it is a quantum statistic problem. This means it can reveal the nature of the system directly by different dimensionalities of the 3D spin lattice and the 2D transfer matrices. 4) It is easy to map the 3D spin-glass Ising model to any other NP-complete problems with two possible states for each element (spin, particle, variable, object, etc.).

The history of studying the knapsack problem (defined as $M_{KP}$) stretches back to Mathews' work on the partition of numbers [16,17]. The knapsack problem has been intensively investigated, since it was determined to be one of the NP-complete optimization problems in 1970s [18,19]. The 0-1 knapsack problem is the simplest among the knapsack problems [20,21]. With the list of a binary decisional variable $x_i$ for $i \in [N]$, of fixed weight ($w_i$) and cost ($c_i$), the aim of the knapsack problem is to maximize the sum of the costs of objects while the sum of the weights of the objects is restricted to be less than or equal to the maximal weight $W_{max}$. The knapsack problem can be used for calculations in the fields of combinatorial mathematics, cryptography, business and so on [22]. Furthermore, the knapsack problem appears in decision-making processes in the different fields, for instance, searching the least cost route for reducing the use of raw materials, selecting the investment portfolio, generating secret key systems, and so on. The knapsack problem also offers many practical applications in various areas, such as project selection, resource distribution, investment decision making, and so on. There are some studies on the relations between the knapsack problem and other NP-complete problems such as K-SAT problem [23,24], traveling salesman problem (TSP) [25,26], etc. The correlation between the knapsack problem and the spin glass model is described briefly as follows: 1) The binary decisional variable in the knapsack problem corresponds to the Ising spin in the spin glass model. An object outside/inside the knapsack corresponds to a spin up/down state. 2) The weighs in the knapsack problem are mapped to the connections (interactions) with randomness in the spin glass model. 3) The classical Hamiltonian corresponding to the knapsack problem can be transformed to a Hamiltonian for the standard all-to-all-connected Ising model with bias terms. Therefore, solving the knapsack problem can be realized by solving in the spin glass model. Maximizing the sum of the costs of the objects in the knapsack problem is equivalent to minimizing the free energy in the spin glass model.

I uncovered the most important feature of the mathematical structures of the 3D Ising models [9, 27,28]. By utilizing the key issue of the nontrivial topological structure, I succeeded in determining the critical point of the ferromagnetic Ising model in a simple cubic lattice to be located at the golden ration ($1/K_c$ = 4.15617384 . . .) [9]. Note a criterion that the critical point of the simple cubic Ising model must be much higher than that ($1/K_c$ = 3.6409569 . . .) of the triangular Ising model [9]. It is worth mentioning recent Monte Carlo simulations [29], in which the critical exponents of the 3D Ising model obtained by taking into account the long-range interactions of spin chains (namely, the nontrivial topological contribution) agree well with my exact solution. Furthermore, the exact solution for the critical exponents [9] agree well with experimental results in various materials as a 3D Ising universality [30–32]. The procedures developed for solving analytically the 3D ferromagnetic Ising models [9,27,28] can be utilized to understand other related ones, but much more complicated problems, such as the 3D spin-glass Ising models [14,15]. The lower bound for the computational complexity of the 3D spin-glass Ising models [14] and the K-SAT problem [15] has been recognized by several groups of computer scientists worldwide [33–35].

My aim of this work is to determine the lower bound of the computational complexity of the knapsack problem $C_L(M_{KP})$. The problem is sketched as follows: In section 2, the lower bound of the computational complexity of the 3D spin-glass Ising model is determined by analyzing the topological structures and understanding the

AMC model. I first inspect the origin of the nontrivial topological structures in the NP-complete problems. I illustrate a phase diagram for the NP vs P problems, in which that there is a NP-intermediate (NPI) problem between the NP-complete problems and the P-problems, while the AMC model is at the border between the NPI problems and the NP-complete problems. This indicates that the AMC model of the 3D spin-glass Ising model cannot collapse directly into the P-problems. In section 3, the lower bound of the computational complexity of the knapsack problem is determined by analyzing the correspondence and the mapping between the knapsack problem and various spin-glass models. The results show the existence of the NPI problems and the AMC model for the knapsack problem. A phase diagram for the NP vs P problems, similar to the spin-glass system, is obtained also for the knapsack problem. Furthermore, attention is given for applications of other NP-complete problems (such as TSP, networks) and comparisons with other optimum algorithms, such as dynamic programming and genetic algorithms, are made. The conclusion is represented in section 4.

## 2. Origin of the nontrivial topological structures in a 3D spin-glass Ising model

**Definition 1.** Let $M_A^D$ be a physical model where the upper script fixes the dimension or named model, and the lower indices indicate the character of the model.

**Definition 2.** Let $C(M_A^D)$ be the computational complexity of the model $M_A^D$.

**Definition 3.** Let $C^U(M_A^D)$ be the upper bound of the computational complexity of $M_A^D$. The upper bound for a model is equal to the computational complexity, as computed by brute force search.

**Definition 4.** Let $C_L(M_A^D)$ be the lower bound of the computational complexity of $M_A^D$.

**Definition 5.** Let $M_{SGI}^{3D}$ denote the 3D spin-glass Ising model, $M_{AMC,SGI}^{3D}$ the AMC model in $M_{SGI}^{3D}$, $M_{SGI}^{SK}$ the Sherrington-Kirkpatrick spin-glass Ising model with all-to-all-connected interactions, and $M_{SGI}^{EA}$ the Edwards-Anderson model with the nearest neighboring interactions.

*2.1. Model establishment*

The Edwards–Anderson model is a model for describing spin-glass systems, in which interactions between only the nearest neighboring spins are considered [4]. Spins are arranged on a 1D, 2D, or 3D lattice with randomly distributed competing interactions. In this section, I focus on the Edwards–Anderson model to show that even with only the nearest neighboring interactions, the long-range spin entanglement exists in the system. In the next section, I also discuss other spin-glass models, for instance, the Sherrington-Kirkpatrick model with the long-range interactions.

The Hamiltonian of a spin-glass Edwards–Anderson model with $S = 1/2$ Ising spins is expressed as [4,14,15]:

$$H = -\sum_{<i,j>} \tilde{J}_{ij} S_i S_j, \tag{1}$$

where the nearest interactions $\tilde{J}_{ij}$ are taken values $\tilde{J}_i$ (i = 1,2,3) along three crystallographic directions. The interactions with different signs are randomly distributed in range of [-J, J] by a Gaussian distribution (or a pseudo-random generator). As usual, I prefer to use the interactions $\widetilde{K}_i = \tilde{J}_i/k_B T$ to replace $\tilde{J}_i$. $k_B$ is the Boltzmann constant and $T$ the temperature. The partition function $\bar{Z}_\alpha$ can be

formulated in a fixed replica α (= 1,2,…R) as [15,27,28]:

$$\bar{Z}_\alpha = (2sinh2\widetilde{K})^{\frac{mnl}{2}} \cdot trace(V_3 V_2 V_1), \quad (2)$$

$$V_1 = \prod_{j=1}^{mnl} exp\{i\widetilde{K}_1^* \cdot C_j\}, \quad (3)$$

$$V_2 = \prod_{j=1}^{mnl} exp\{i\widetilde{K}_2 s'_j s'_{j+1}\}, \quad (4)$$

$$V_3 = \prod_{j=1}^{mnl} exp\{i\widetilde{K}_3 s'_j s'_{j+mn}\}. \quad (5)$$

Here, m, n, and l are the numbers of the lattice sizes along three crystallographic directions of the Ising spin system. $\widetilde{K}_1^*$ is defined by [15,27,28],

$$tanh\widetilde{K}_1^* \equiv e^{-2\widetilde{K}_1}. \quad (6)$$

The matrices $C_j$ and $s'_j$ are defined as the direct products of Pauli matrices $\sigma_j$ (j = 1,2,3) [15].

In a spin-glass 3D Ising model with randomly distributed positive and negative interactions between spins, containing frustration, different ground states, such as a ferromagnetic state, an antiferromagnetic state or a spin glass state may occur. In a limit case, if all the interactions are ferromagnetic and randomly distributed, a random ferromagnetic state without frustration may occur. In this work, I am interested only in the spin-glass 3D Ising lattice with strongly competing interactions in general cases (or worst cases for computational complexity), in which, in the presence of frustration, neither ferromagnetic nor antiferromagnetic interaction is dominant so that it is not easy to determine the ground state of the system to be ferromagnetic, antiferromagnetic, or spin glass.

The partition function Z of the spin-glass system can be calculated from the product of the partition functions $\bar{Z}_\alpha$ for all fixed replicas (α = 1,2,..,R),

$$Z = \prod_{\alpha=1}^{R} \bar{Z}_\alpha. \tag{7}$$

Thus, it is enough to focus on $\bar{Z}_\alpha$ for the lower bound of the computational complexity. The elements in the transfer matrices and the partition function are calculated with a combination of different states of spins on the 3D lattice points.

*2.2.   Non trivial topology analysis*

The nonlocal effect in the partition function (2) (see also the transfer matrix (5)) of the 3D Ising model originates from the contradictory between the 3D character of the lattice and the 2D character of the transfer matrices used in the quantum statistics mechanism. In what follows, I inspect the origin of the nonlocal effect:

In a 3D Ising model with the lattice size $N = mnl$, Ising spins are assigned on every lattice point. The numbers $(i, r, s)$ denote lattice points running from $(1, 1, 1)$ to $(m, n, l)$ along three crystallographic directions in the 3D lattice. One may denote the lattice points, a layer by a layer, by the number $j = [mn(s-1) + m(r-1) + i]$, which runs in a sequence as $1, 2, 3, \ldots, mnl$. Such two representations are equivalent. The size of the transfer matrices in spinor representation for the 3D Ising model is $2^N \times 2^N$. It is noticed that the sequence in a process of a layer by a layer is the simplest one for mapping a 3D lattice into a 2D "lattice" (a matrix) for the quantum states described by the transfer matrices. For studying the 3D Ising model, this sequence can remain some basic characters of the 2D Ising model, and make the procedure for solving the 3D Ising model as simple as possible. It was understood that any other sequences will make the problems much more complicated [14].

Let us run $j$. For the first layer ($s = 1$), corresponding to the running from $(1,1,1)$ to $(m, n, 1)$, we have $j = 1, 2, 3, \ldots, m$ for the first line ($r = 1$), $j = m+1, m+2, m+3, \ldots, 2m$ for the second line ($r = 2$), …, and $j = (n-1)m+1, (n-1)m+2, \ldots, mn$ for the last line ($r = n$).

For the second layer ($s = 2$), corresponding to (1, 1, 2) to ($m$, $n$, 2), $j$ runs from ($mn$+1), ($mn$+2), ($mn$+3),..., $2mn$. It runs in all the way to the last layer ($s = l$), $j$ runs from ($mn$($l$-1)+1), ($mn$($l$-1)+2), ($mn$($l$-1)+3),..., $mnl$. The first spin (1,1,1) in the first layer and the first spin (1, 1, 2) in the second layer correspond to $j = 1$ and $j = (mn+1)$, respectively. Figure 1 illustrates a spin-glass Ising model on a 3D lattice with the size of 3×3×3, for example, which is mapped into the spin arrangement on a 2D lattice with the size of (3×3+3×3+3×3), as arranged in the transfer matrix. The spins assigned on the lattice are randomly distributed, pointing up or down. Green, purple, and blue colors represent the interactions along three crystallographic directions, which can be randomly distributed ferromagnetic or antiferromagnetic. In the transfer matrix, the interactions with blue color show the crossings. It is emphasized here that the interaction between the two nearest neighboring spins behaves as a long-range interaction, which involves the entanglements of all the spins in a plane. This is the origin of the nontrivial topological structures.

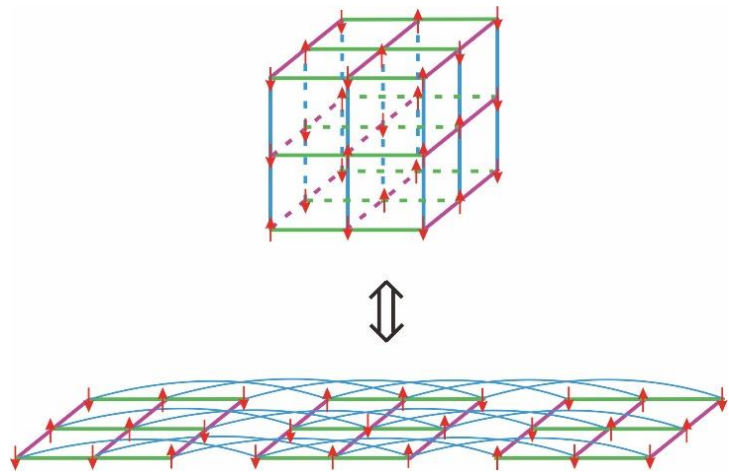

**Figure 1.** Illustration of a spin-glass Ising model on a 3×3×3 lattice, for example, which is mapped into the 2D spin arrangement on a (3×3+3×3+3×3) lattice, as in the transfer matrix. The spins at the bottom,

middle, and top layers of the 3D Ising lattice are mapped to those at the left, middle, and right 2D Ising lattices, respectively. The spins pointing up or down are assigned on the lattice and randomly distributed. Green, purple, and blue colors represent the interactions along three crystallographic directions. In the transfer matrix, the interactions with blue color show the crossings, indicating the existence of non-trivial topological structures in the 3D spin-glass Ising model.

For the fully ferromagnetic case, for simplicity, we can apply the cylindrical crystal model preferred by Onsager [8], in which we wrap our crystal on cylinders. However, unlike in the solid energy band theory for one-electron approximation in which the periodic boundary conditions can be applied along three crystallographic directions, we can perform the periodic boundary condition only along one crystallographic direction in the present system with many-body spin-spin interactions. After performing the periodic boundary condition, the running number $j$ can be reduced to $j = [(s-1)n + r]$, running as $j = 1,2,3,\ldots,nl$ in a plane. The size of the transfer matrices in spin or representation for the 3D ferromagnetic Ising model is reduced to be $2^{nl} \times 2^{nl}$. For the spin glass case, even such a periodic boundary condition cannot be employed, mainly due to the randomness of interactions. For the 3D spin-glass Ising model in a fixed replica, the size of the transfer matrices in spinor representation remains $2^N \times 2^N$. Of course, the nonlocal effects are the natural character of the 3D many-body spin-spin interacting models in the quantum statistics mechanism. The boundary factors for 2D (or 3D) models can be neglected in the thermodynamic limit, whereas the internal factors in the transfer matrix (5) for 3D cases cannot be neglected, since they appear at each lattice point. Indeed, the nonlinear terms of $s'_j s'_{j+mn}$ in the transfer matrix $V_3$

(see Eq (5)) indicate the existence of the nontrivial topological structures. All these characters together with random interactions and spin alignments (and also frustrations) cause the system to be NP-complete [2,3,5].

*2.3.    Absolute minimum core model*

A 3D spin-glass Ising lattice can be constructed by stacking $l$ layers of 2D spin-glass Ising lattices. This is the simplest way to construct the 3D spin-glass Ising model layer by layer, while keeping the characters and (thus the physical properties) of the 3D (and also 2D) spin-glass Ising model. Other ways of constructions may cause much more complicated procedures (referred to Theorem 2 in [14]). According to Theorem 2 in [14], to find the exact solution of the 3D spin-glass Ising model, any algorithms cannot break the global effects of entanglements in the AMC model. Figure 2 shows an example for the AMC model $M_{AMC,SGI}^{3D}$. Notice that two layers are needed to represent the AMC model, in which the solid lines represent the bottom ($l = 1$) layer with the intralayer interactions and the interlayer interaction between the two layers, while the dashed lines show that there are no intralayer interactions on the top layer ($l = 2$). Spins (red arrows) at every lattice point of a two-level grid lattice align along with randomly distributed directions, caused by randomly distributed interactions, while spins (blue double arrows) in some plaquettes show the frustrations in the spin-glass system. It has been proven [14,15] that for the 3D spin-glass Ising model, the upper bound of the complexity (by brute force search) of the AMC model gives its lower bound. That is,

$$C_L(M_{SGI}^{3D}) \geq C^U(M_{AMC,SGI}^{3D}). \tag{8}$$

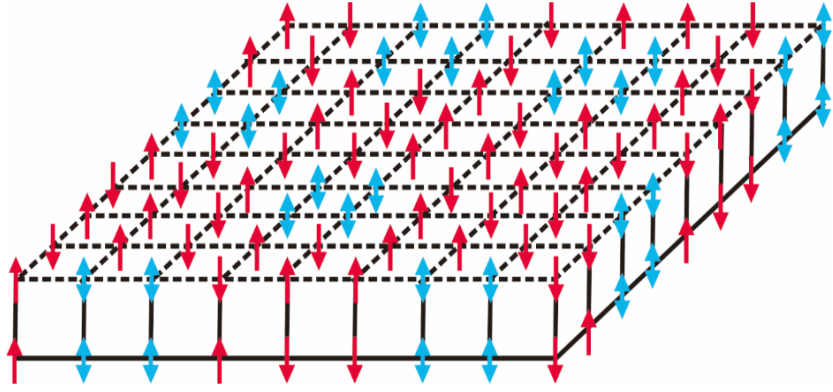

**Figure 2.** Schematic illustration of an AMC model, $M^{3D}_{AMC,SGI}$, for the 3D spin-glass Ising model, in which spins (red arrows) at every lattice point of a two-level grid lattice (with the lattice size $N = mnl$, here $m = n = 9$ and $l = 2$ as an example) align along with randomly distributed directions, caused by randomly distributed interactions between spins. Moreover, spins in some plaquettes are represented by blue double arrows, to show the existence of frustrations in the spin-glass system.

### 2.4. NP-intermediate problems

Ladner [36] showed that, assuming P ≠NP, there exist NPI problems, that is, problems in NP that are neither in P nor NP-complete. Ladner explicitly constructed NPI problems by removing strings of certain lengths from NP-complete languages via a diagonalization technique that is colloquially known as blowing holes in problems [36,37].

I am interested in whether there is a NPI area between the NP-complete problem (the AMC is located on its border) and the P-problem in our system. In other words, if the AMC model can collapse directly into the P-problem? We have determined the lower bound of the complexity of the 3D spin-glass Ising model is $O(2^{nm})$, which is

subexponential time [14,15]. We have [14,15]

$$O(2^{nm}) = O(2^{N^{2/3}}) = O((1+\varepsilon)^N), \qquad (9)$$

when $n = m = N^{1/3}$, and with $\varepsilon \to 0$ and $\varepsilon \neq 1/N$. It is known that there are some quasi-polynomial times, for instance, $O(N^{lgN})$, $O(N^{lglgN})$, etc. Thus, we have

$$O(2^N) \gg O(2^{nm}) = O((1+\varepsilon)^N) = O\left(2^{N^{\frac{2}{3}}}\right) \gg O(N^{lgN}) \gg O(N^{lglgN}) \gg O(N^P).$$

(10)

Figure 3 illustrates schematically these results obtained above to be a phase diagram of different complexities. The lower bound of the complexity of the 3D spin-glass Ising model is at the boundary of the area for the NP-complete problems. There is a NPI area between the NP-complete problems (the AMC model is on its border) and the P-problems. The NPI problems have the computational complexity (including quasi-polynomial times $O(N^{lgN})$, $O(N^{lglgN})$, etc.) less than $O\left(2^{N^{2/3}}\right)$ and larger than $O(N^P)$. The NPI problem may be constructed by removing some interactions and/or spins in the AMC model, following the Ladner's process [36]. Namely, we have some incomplete AMC models as the NPI problems. It is concluded that the AMC model cannot collapse directly into the P-problem.

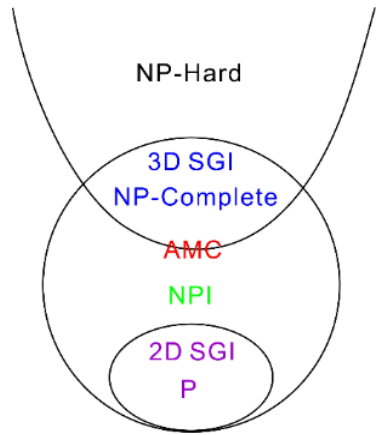

**Figure 3.** Phase diagram for the 3D spin-glass Ising model. In the phase diagram, 3D SGI represents NP-complete problems, and P represents polynomial problems (2D SGI). NPI exists between NP-complete and P problems, while AMC is located on the border of NP-complete and NPI regions.

*2.5. Strategy for an optimum algorithm*

From the analysis above, I propose the following strategy for developing an optimum algorithm for calculations of physical properties (such as, the ground state, the free energy, the critical point, the phase transitions, and the critical phenomena) of the 3D spin-glass Ising model.

1) Fix z-layers (z = 1, 2, 3, …) of the AMC model as an element of the algorithms, while performing a parallel computation of $l/z$ layers of this element.
2) Compare the precision as well as the accuracy of the results obtained by the above procedures, and determine the optimum value of z.

In this way, one can design the optimum algorithm to find/reach the exact solution with the sufficient accuracy and within the high precision in the shortest time. It can be improved greatly from the present status of $O(1.3^N)$ [6] to $O((1 + \varepsilon)^N)$ with $\varepsilon \to 0$ and $\varepsilon \neq 1/N$ [14,15], the best case if one can succeeded in the optimum value z = 1. Since the 3D spin-glass Ising model is catalogued to NP-complete set, the optimum algorithm can be employed to compute the properties of other NP-complete problems (for instance, TSP, K-SAT problem, knapsack problem, neural networks, etc.).

**3. Lower bound of the computational complexity of the knapsack problems**

## 3.1. Definitions of the knapsack problem

**Definition 6.** Let $M_{KP}$ denote the knapsack problem, $M_{KP}^{all}$ the knapsack problem on an all−to−all−connected network, $M_{KP}^{3D}$ the 3D knapsack problem, $M_{KP}^{2D}$ the 2D knapsack problem, and $M_{AMC,KP}^{3D}$ the AMC model in the 3D knapsack problem.

**Definition 7.** Let $M_{TSP}$ denote the TSP, $M_{TSP}^{3D}$ the 3D TSP, $M_{l=2,TSP}^{3D}$ the TSP model on a two-level grid lattice, $M_{TSP}^{2D}$ the TSP model on a 2D lattice, and $M_{AMC,TSP}^{3D}$ the AMC model in the 3D TSP.

I consider the 0-1 knapsack problem, which is the simplest one among all the knapsack problems [18–21]. Having the list of a binary decisional variable $x_i$ for $i \in [N]$, of fixed weight ($w_i$) and cost ($c_i$). The goal of the knapsack problem is to maximize the sum of the costs of the objects while the sum of the weights of objects is restricted to be less than or equal to the maximal weight $W_{max}$. This problem is NP-complete classically, and can be written as [18–21]:

$$\max \sum_{i \in [N]} c_i x_i$$

$$s.t. \sum_{i \in [N]} w_i x_i \leq W_{max}, \quad x_i \in \{0,1\} \tag{11}$$

Here, an object index $i$ runs from 1 to $N$, weights correspond to integer-valued numbers. For the 0-1 knapsack problem, a binary variable $s_i$ is introduced for $x_i \in \{0,1\}$. $s_i$ is equal to 1 when an object is inside the box and 0 otherwise (see Figure 4(a)). For the Ising spin model, a binary variable $s_i$ is introduced for $x_i \in \{-1,1\}$. $s_i = \pm 1$ represents the spin up and the spin down, respectively (see Figures 4(b–d)). The total weight and total value then read

$$W = \sum_{i=1}^{N} w_i s_i \tag{12}$$

and

$$C = \sum_{i=1}^{N} c_i s_i, \quad (13)$$

respectively. I further introduce auxiliary binary variables $a_j$, where the index $j$ runs from 1 to $W_{max}$. The formulation above for the 0-1 knapsack problem looks simple, but the time needed for the computing procedure increases rapidly with the size of the system, as other NP-complete problems.

For the 0-1 knapsack problem, the problem of placing the first $i$ items of objects into a knapsack with the maximal weight $W_{max}$ (or capacity v), if I focus only on the strategy of the $i$th object (outside or inside the knapsack), then it can be transformed into a problem involving only the first $i$-1 items of objects (see the classical dynamic programming (DP) algorithm proposed by Bellman [38]. There are two cases for the $i$th object: If the $i$th object is outside the knapsack, the problem will become the first $i$-1 items of objects are put inside the knapsack with capacity v and the cost $f_{i-1}[v]$. If the $i$th object is inside the knapsack, the problem will become the first $i$-1 items of objects are put inside the knapsack with capacity v-$w_i$ and the cost $f_{i-1}[v-w_i]$ plus the cost $c_i$ for putting the $i$th object inside the knapsack. The process of the 0-1 knapsack problem can be mapped to the problem of Ising spins, in which the states of the $i$th spin is entangled with the states of the first $i$-1 spins in the lattice. Note that putting the objects outside or inside the knapsack corresponds to up or down alignment of the spins, while the weighs are transformed to the interactions between spins. In principle, determining the ground state of a N-spin system equalizes to determining the states of the N-th spin that entangled with all the first N-1 spins in the system. My purpose of this work is to find the lower bound of the computational complexity of the knapsack problems.

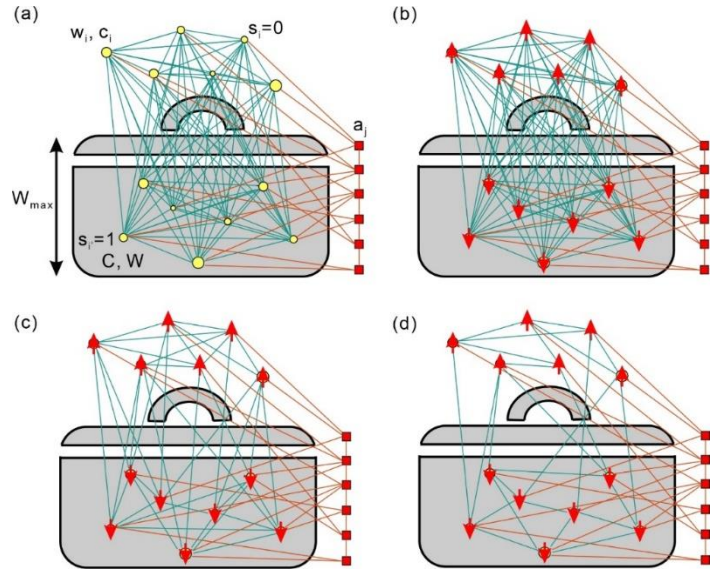

**Figure 4.** Sketch of (a) the knapsack problem [39], which corresponds to (b) an all-to-all-connected Ising network, (c) an Ising network with six connections (interactions) on a spin, and (d) an Ising network with four connections (interactions) on a spin. Different items of weight $w_i$ and cost $c_i$ can be placed in the suitcase (assign binary variable $s = 1$) or left outside ($s = 0$). Compared with Figure 8 in [39], I have added up and down spins on these items for indicating the states of inside or outside the suitcase. The maximal weight of the suitcase is bounded by $W_{max}$. The solution can be obtained by searching the energy minimum for all the configurations for combined item ($s_i$) and auxiliary ($a_i$) spins, combined into an Ising network.

### 3.2. Overview of research on the complexity of knapsack problems

In this subsection, I present a brief overview of the history and current status of research on the complexity of knapsack problems. At first, I mention the earliest work on this subject: The knapsack problem has been known since over a century, which can be stretched back to Mathews' work on the partition of numbers [16,17]. The first algorithmic studies were published in the Fifties [40,41], an intense research activity started in the Sixties. The knapsack problem has been intensively investigated, since it was determined to be one of NP-complete optimization problems in Seventies [11,13,18]. In general, all optimization problems are NP-hard. While the ''simplest'' single knapsack problems are NP-hard in the weak sense, i.e., they may be solved in pseudopolynomial time through dynamic programming, most variants and generalizations are NP-hard in the strong sense (i.e., they cannot be solved by pseudo-polynomial time algorithms unless P = NP).

For a detailed overview of recent advances on the knapsack problems, readers can refer to Cacchiani, et al.'s survey [42,43]. Part I of this survey [42] covered the classical single knapsack problems and their many variants and generalizations: Subset sum, item types, setup, multiple-choice, conflict graphs, precedences, sharing, bilevel, robust, among others. It also focused on extensions and generalizations provides pointers to the variants that are especially attractive from the point of view of possible future investigations. Moreover, Part II [43] was mainly devoted to multiple, multidimensional (vector and geometric), and quadratic knapsack problems, but also contained a succinct treatment of online and multiobjective knapsack problems.

*3.3.     Relation between knapsack problems and spin-glass models*

Solving optimization problems is highly demanded in various fields of science ranging from physics to biology to finances, and to information technologies [44,45]. The relation between the knapsack problem, the spin glass and the Ising machine were discussed in [39]. In what follows, I determine the lower bound of the complexity of the knapsack problems via set up the correspondence between the two problems (namely, spin-glass and knapsack).

The classical Hamiltonian corresponding to the knapsack problem then reads [39]:

$$H = \alpha\left(1 - \sum_{j=1}^{W_{max}} a_j\right)^2 + \alpha\left(\sum_{j=1}^{W_{max}} ja_j - \sum_{i=1}^{N} w_i s_i\right)^2 - \beta \sum_{i=1}^{N} c_i s_i$$

(14)

where $\alpha$ and $\beta$ are parameters for the simulation, chosen to ensure that the solution is the global minimum of the Hamiltonian (14). In the Hamiltonian (14), the weight $w_i$ is randomly distributed. The Hamiltonian (14) can be rewritten as the standard all-to-all-connected Ising model with bias terms $h_n$ as [39]:

$$H = -\sum_{n<m}^{N+W_{max}} \tilde{J}_{nm} s_n s_m - \sum_{n=1}^{N+W_{max}} h_n s_n, \qquad (15)$$

where the Ising spin $s_n = \pm 1$, The equivalence between (14) and (15) can be clarified as follows: $\tilde{J}_{nm}$ denotes the Ising coupling matrix formed by weights, and $h_n$ is an effective magnetic field formed by the combination of cost and weight. The parameters are inferred from the original encoding given in Eq (14). In the Hamiltonian (15), the interactions $\tilde{J}_{nm}$ of different signs are randomly distributed in range of [-J, J]. The random distribution of the interactions in the Hamiltonian (15) is transformed from the randomly distributed weight $w_i$ in the Hamiltonian (14), which provides the possibility of existing plaquettes containing frustrations in the spin-glass system as illustrated in Figure 2. To clarify a detailed correspondence between (14) and (15), the terms with parameters $\alpha$ and $\beta$ in (14) are transformed into the interaction terms $s_n s_m$ and the terms with the effective magnetic field $h_n$ in (15), respectively. The variable $s_i$ in the Hamiltonian (14) takes two values 1 and 0, while the Ising spin in the Hamiltonian (15) takes two values, 1 and -1. The interaction terms $s_n s_m$ in (15) have four combinations (++, +-, -+, --), resulting in two values, 1 and -1. Thus, one can carry out a mapping from the two values 1 and 0 for the variable $s_i$ in (14) into the two values 1 and -1 for the interaction terms $s_n s_m$ in (15). This indicates clearly the correspondence between the terms $w_i s_i$ in (14) and the terms $\tilde{J}_{nm} s_n s_m$ in (15).

Therefore, the interaction couplings $\tilde{J}_{nm}$, in (15) are transformed directly from the weights $w_i$ in (14). On the other hand, it is easy to see the connection between the terms $c_i s_i$ in (14) and the terms $h_n s_n$ in (15).

The knapsack problem can be mapped into the spin-glass all-to-all-connected Ising model with appropriate parametric correspondence (see Figure 4(b)). As mentioned, an object is inside or outside the box in the knapsack, corresponds to a spin pointing up or down in the spin-glass Ising model. As mentioned in the last section, I am interested only in the worst cases for computational complexity of the spin-glass 3D Ising model with strongly competing interactions in the presence of frustration, while searching the ground state among all the possible states (including ferromagnetic, antiferromagnetic, and spin–glass states).

*3.4.    Lower bound of the computational complexity of the knapsack problem*

**Theorem 1.** The lower bound of the computational complexity of the knapsack problem in the 3D lattice, $C_L(M_{KP}^{3D})$ or $C^U(M_{AMC,KP}^{3D})$, is in subexponential and superpolynomial.

*Proof of Theorem 1.* The standard all-to-all-connected spin-glass Ising model is the so-called Sherrington-Kirkpatrick model [46,47], $M_{SGI}^{SK}$, in which the <i, j> sum in the Hamiltonian is over all bonds. It is an Ising model with long-range ferromagnetic as well as antiferromagnetic couplings for frustrated states. Ising spins interact through infinite-ranged exchange interactions, which are independently distributed with a Gaussian probability density. Sherrington and Kirkpatrick observed that it is an exactly solvable model of a spin glass in the limit of infinite interactions, within the mean-field theory. However, for a brute force search of the ground state of the Sherrington-Kirkpatrick model [46,47], the job is NP-complete for computer. The Sherrington-

Kirkpatrick model can be reduced to a 3D spin-glass Ising (Edwards-Anderson) model with the nearest neighboring interactions $M_{SGI}^{EA}$, by simplify neglecting the long-range interactions between spins.

At first, the Sherrington-Kirkpatrick model $M_{SGI}^{SK}$ can be reduced to the Edwards-Anderson model only with the nearest neighboring interactions, $M_{SGI}^{EA}$, by cutting the long-range interactions between spins. Second, it can be reduced further by reducing the nearest neighboring interactions to be six. In such an artificial lattice with six connections (interactions) on a spin, the averaged number for the interactions per spin is actually three, since one bond (interaction) connects two spins. Such a spin-glass system can be arranged either in a 3D lattice or a 2D lattice. If the spin-glass Ising network with three interactions per spin can be assigned only on a 3D lattice, $M_{SGI}^{3D}$, it will be NP-complete. If it can be assigned on a 2D triangle (honeycomb) lattice without crossings, $M_{SGI}^{2D}$, it will be a P-problem. The spin-glass Ising network with four connections on a spin (i.e., two interactions per spin) is a P-problem, since it can be assigned on a 2D rectangular (or square) lattice without crossings.

Although the Sherrington-Kirkpatrick model with the long-range interactions can be solved analytically by a mean-field approach [46,47], for computer simulations, one has to summarize all the interactions (including all the connections) in the system. Thus, the computational complexity of the Sherrington-Kirkpatrick model should be much larger than that of the Edwards-Anderson model with the nearest interactions only. The following relations are held for spin-glass systems:

$$C(M_{SGI}^{3D,SK}) > C(M_{SGI}^{3D,EA}) \geq C(M_{AMC,SGI}^{3D}) \gg C(M_{SGI}^{2D,EA}). \tag{16}$$

Therefore, the lower bound of the computational complexity of the 3D Edwards-Anderson model $C_L(M_{SGI}^{3D,EA})$ is lower than that of the Sherrington-Kirkpatrick model

$C_L(M_{SGI}^{3D,SK})$. That is,

$$C_L(M_{SGI}^{3D,EA}) < C_L(M_{SGI}^{3D,SK}). \qquad (17)$$

Thus, it is good enough to figure out the former. Moreover, we can use the results obtained in section 2 for the 3D spin-glass Ising model.

Utilizing the relationship between the knapsack problems and the Ising model, we assure that an AMC model exists also in the knapsack problem. Corresponding to Eq (16) for the spin-glass systems, the following relations are held for the knapsack problems:

$$C(M_{KP}^{all}) > C(M_{KP}^{3D}) \geq C(M_{AMC,KP}^{3D}) \gg C(M_{KP}^{2D}) \qquad (18)$$

A phase diagram for the NP vs P problems is illustrated in Figure 5 for the knapsack problems. Accordingly, the lower bound of the complexity of the knapsack problem $C_L(M_{KP}^{3D})$ is that as calculated by brute force search of the AMC model, $C^U(M_{AMC,KP}^{3D})$. Namely, similar to Eq (8),

$$C_L(M_{KP}^{3D}) \geq C^U(M_{AMC,KP}^{3D}). \qquad (19)$$

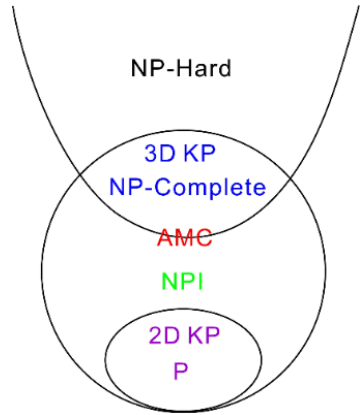

**Figure 5.** Phase diagram for the 0-1 knapsack problem. In the phase

diagram, 3D KP represents the NP-complete problems, and P represents polynomial problems (2D KP). NPI exists between NP-complete and P problems, while AMC is on the border of NP-complete and NPI regions.

On the other hand, to find the most efficient algorithms for solving the knapsack problem, I may need to arrange the knapsacks (like spins) on a 3D lattice. By adjusting/removing the unimportant weights in the knapsacks, one may disconnect some long-range interactions between spins, to obtain an "easier" arrangement of spins with only the nearest interaction in the 3D lattice. Like in a 3D spin-glass Ising lattice, all the knapsacks can be constructed by stacking $l$ layers of the knapsacks located on 2D lattices. This is the simplest way and thus the optimum algorithms to construct the knapsacks on a 3D model layer by layer, while keeping the characters and (thus the physical properties) of the knapsacks. Other ways of constructions may cause much more complicated procedures (referred to Theorem 2 in [14]). I have

$$C_L(M_{KP}^{3D}) = C_L(M_{SGI}^{3D}) \geq C^U(M_{AMC,SGI}^{3D}) = C^U(M_{AMC,KP}^{3D}). \tag{20}$$

As revealed in [14,15] and in the last section, the computational complexities $C^U(M_{AMC,SGI}^{3D})$ and $C_L(M_{SGI}^{3D})$ are in $O((1+\varepsilon)^N)$ with $\varepsilon \to 0$ and $\varepsilon \neq 1/N$. Therefore, the computational complexities $C^U(M_{AMC,KP}^{3D})$ and $C_L(M_{KP}^{3D})$ are in the same class, which are subexponential and superpolynomial.

Similar to the procedure proposed in the last section, taking z-layers of the AMC models as an element of the algorithms, we may develop an algorithm for performing a parallel computation of $l/z$ layers of the AMC models for the knapsack problems. In this way, I can succeed in designing the optimum algorithm to find the exact solution with the sufficient accuracy and within the high precision in the shortest time. It can be improved greatly from the present status of $O(1.3^N)$ [6] to $O((1+\varepsilon)^N)$ with $\varepsilon \to 0$ and $\varepsilon \neq 1/N$ [14,15].

*3.5. NP-intermediate problems*

**Theorem 2.** A NPI area exists between the NP-complete problems and the P-problems for the knapsack problem.

*Proof of Theorem 2.* According to the results in the last section (see also Figure 3), there exists a NPI problem $M_{NPI,SGI}$ for spin-glass Ising models, which is in between $M_{SGI}^{3D}$ and $M_{SGI}^{2D}$, while $M_{AMC,SGI}^{3D}$ is the border between $M_{SGI}^{3D}$ and $M_{NPI,SGI}$. Similarly, as illustrated in Figure 5 for phase diagram for the knapsack problem, a NPI problem $M_{NPI,KP}$ exists for the knapsack problem, which is located in between the NP-complete problem $M_{KP}^{3D}$ and the P-problem $M_{KP}^{2D}$ and thus the AMC model $M_{AMC,KP}^{3D}$ is the border between $M_{KP}^{3D}$ and $M_{NPI,KP}$.

The 0-1 knapsack problem is the most basic problem among all the knapsack problems, which consists of the designed states and the basic concepts of equations. Many others are treated as its generalization and can be transformed to the 0-1 knapsack problem. Therefore, the results obtained above for the 0-1 knapsack problem can be applied for them.

*3.6. Applications for other NP-complete problems*

It is worth noting that this study can be extended to other NP-problems, such as K-SAT problem [15,23,24], TSP [25,26], neural networks [48,49], etc. In particular, in recent years, the neural networks have been applied in rapidly progressed fields of deep learning, artificial intelligence, and so on. It is very visual that the networks illustrated in Figure 4 for the knapsack problem as well as the Ising models can be transformed into the neural networks. The conventional approach to these problems is to study the complexity of an equivalent yes/no question. In the following, we take the TSP [25,26] as an example, which is defined as follows.

A traveling salesman has to visit all N cites and return to the starting point at the end of the tour (also called Chinese Postman's problem [50]). Taking into account the two traversals (in opposite directions) of each tour and the arbitrariness of the starting city, there are (N-1)!/2 distinct tours. The TSP is asking to find the shortest tour(s) (the optimal one) among them, which can be described also in the following problem: Given a graph G with costs on the edges, find a cycle in G that visits every node exactly once and minimizes the length of the cycle. This problem is converted to the question: Given a graph G and an integer k, does G have a TSP tour of cost at most k? Although this transformation loses some of the structure of the original problem, it captures the essential difficulty of the TSP problem because we can solve the original problem by using the yes/no question as a subroutine. Upon the dimensionality (namely, 2D or 3D) of the tours in the TSP, it can be catalogued to a NP-complete problem or a P problem. With a similar procedure to this work, we can find the AMC model and the NPI problem for the TSP. Figure 6 illustrates the TSP model on a two-level grid lattice with a small size (with the lattice size $N = mnl$, here $m = n = 9$ and $l = 2$), which is NP-complete. The AMC model for the TSP identifies to the difference between a two-level ($l = 2$) grid TSP model and a 2D TSP model, namely, $M_{AMC,TSP}^{3D} = M_{l=2,TSP}^{3D} - M_{TSP}^{2D}$, which is NP-complete also. Clearly, either the 3D spin-glass Ising model or the knapsack problem can be transformed into the TSP, the K-SAT problem and neural networks and so on. Even some information might be lost during the transformation, the essential difficulty remains, and thus the lower bound of the computational complexity maintains the same. This means that the lower bound of the computational complexity of all the NP-complete problems is in the same universality class (being superpolynomial and subexponential).

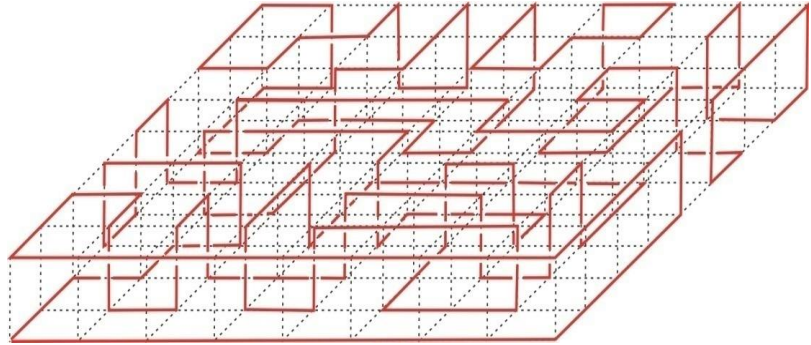

**Figure 6.** Schematic illustration of a TSP model on a two-level grid lattice (with the lattice size $N = mnl$, here $m = n = 9$ and $l = 2$), $M^{3D}_{l=2,TSP}$. The black dashed lines represent the lattice, while the red solid lines represent the tour. Here, I illustrate a tour as an example to connect all the lattice points (cities) in the two-layers ($l = 2$). There exist some crossings in the tour, which represent the character of the 3D space. In order to illustrate the connections, some solid lines are drawn to be not fitted with the dashed lines for the two-level grid lattice. The AMC model for the TSP identifies to the difference between a two-level ($l = 2$) grid TSP model and a 2D TSP model.

Similar to the 3D spin-glass Ising model (Figure 3) and knapsack problem (Figure 5), a phase diagram for the TSP is illustrated in Figure 7, in which NPI region exists between NP-complete and P problems, while the AMC model is on the border of NP-complete and NPI regions.

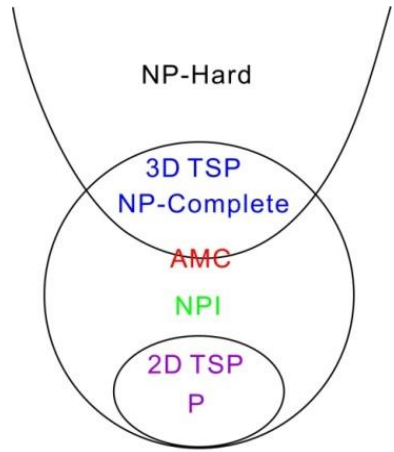

**Figure 7.** Phase diagram for the TSP. In the phase diagram, 3D TSP represents the NP-complete problems, and P represents polynomial problems (2D TSP). NPI exists between NP-complete and P problems, while AMC is on the border of NP-complete and NPI regions.

*3.7. Comparison with other optimum algorithms*

In this subsection, I compare the optimum algorithm suggested in subsection 2.5 with other optimum algorithms, such as dynamic programming and genetic algorithms.

The dynamic programming has been developed to investigate different NP-complete problems, such as K-SAT [51], knapsack problem [19,52,53], TSP [54], etc. Bertsimas and Demir presented an approximate dynamic programming approach for the multidimensional knapsack problem [52]. Woeginger discussed whether a dynamic programming formulation guarantees the existence of a fully polynomial time approximation scheme [53]. Pisinger gave an overview of some exact solution approaches and to show that the knapsack problem is difficult to solve for these algorithms including the dynamic programming algorithms for a variety of test problems [19]. Martello et al. gave an overview of the techniques for solving hard knapsack problems, with special emphasis on the addition of cardinality constraints, dynamic programming, and rudimentary divisibility [18].

The genetic algorithm has been developed to study various NP-complete problems, such as K-SAT [55], knapsack problem [56], spin glass models [57], TSP [58,59], etc. Melkman and Akutsu [55] showed that the general case can be solved in $O(1.871^n)$ time for studying the problem of finding a singleton attractor of a Boolean network consisting of n nested canalyzing functions. Chu and Beasley presented a heuristic based upon genetic algorithms for the multidimensional knapsack problem [56]. Large numbers of ground states of the three-dimensional $\pm J$ random-bond Ising model were calculated by using a combination of a genetic algorithm and cluster-exact approximation [57]. Snyder and Daskin presented an effective heuristic for the generalized TSP, which combines a genetic algorithm with a local tour improvement heuristic [58]. Larranaga et al. gave a review of the different attempts made to solve the TSP with genetic algorithms [59], and presented crossover and mutation operators, developed to tackle the TSP with genetic algorithms with different representations such as: Binary representation, path representation, adjacency representation, ordinal representation, and matrix representation.

Although the dynamic programming and the genetic algorithms are very efficient algorithms for studying the NP-complete problems with short time, and some researchers claimed the single knapsack problems (NP-hard in the weak sense) may be solved in pseudopolynomial time through dynamic programming, these algorithms have disadvantages as follows: These algorithms must take some approximates [52,53,57] or with some particular constraints or by a heuristic approach. Actually, they did not realize finding an exact solution for the NP-complete problems for large size scale, because they did not determine the basic character of the NP-complete problems. In order to derive the exact solution of the NP-complete problems, any algorithms must calculate all the states of the AMC model by brute force search. On the other hand, the optimum algorithm proposed in subsection 2.5 can find the exact solution of the NP-complete problem in subexponential time, because it determines the basic element (i.e., the AMC model) of the NP-complete problems.

## 4. Conclusions

In conclusion, I inspected the origin of the nontrivial topological structures and confirmed the existence of the AMC model in the knapsack problems. I proved that the NPI problems exist between the NP-complete problem and P-problems, while the AMC model is at the border between the NPI and the NP-complete problems. The AMC model of the knapsack problem cannot collapse directly into the P-problem. I determined the lower bound of the computational complexity of the knapsack problems $C_L(M_{KP})$, being in subexponential and superpolynomial. Under the guide of the results, one may develop the optimum algorithms, within a framework of a parallel computation of $l/z$ layers of the z-layer AMC models to solve combinatorial optimization problems in the shortest time (might be improved greatly from $O(1.3^N)$ to $O((1 + \varepsilon)^N)$ with $\varepsilon \to 0$ and $\varepsilon \neq 1/N$). The strategy proposed in this work for developing an optimum algorithm can be applied to compute the properties of other NP-complete problems, such as TSP and neural networks. This work sheds a light on complexity theories for various fields of science ranging from physics to biology to finances, and to information technologies.

**Use of Generative-AI tools declaration**

The author declares he has not used Artificial Intelligence (AI) tools in the creation of this article.

**Acknowledgments**

This work has been supported by the National Natural Science Foundation of China under grant number 52031014.

**Conflict of interest**

The author declares no conflict of interest.